\documentclass[conference, letterpaper]{IEEEtran}
\usepackage[letterpaper, top=1.92cm, bottom=3.2cm, left=1.92cm, right=1.92cm]{geometry}
\usepackage{amsmath,amsfonts}

\usepackage{algorithm}
\usepackage{algpseudocode}
\usepackage{array}
\usepackage[caption=false,font=normalsize,labelfont=sf,textfont=sf]{subfig}
\usepackage{textcomp}
\usepackage{stfloats}
\usepackage{url}
\usepackage{verbatim}
\usepackage{color}
\usepackage{graphicx}
\usepackage{mathtools}
\usepackage{cite}
\usepackage{bm}

\begin{document}

\title{Denoising and Augmentation: A Dual Use of Diffusion  Model for Enhanced CSI Recovery\\
}


\author{\IEEEauthorblockN{Yupeng Li\textsuperscript{\dag}, Ruhao Zhang\textsuperscript{\ddag}, Yitong Liu\textsuperscript{\ddag}, Chunju Shao\textsuperscript{\dag}, Jing Jin\textsuperscript{\dag}, Shijian Gao\textsuperscript{*},}
 	\IEEEauthorblockA{
\textsuperscript{\dag}China Mobile Research Institute, Beijing, China \\
\textsuperscript{\ddag}Beijing University of Posts and Telecommunications, Beijing, China\\
\textsuperscript{*}The Hong Kong University of Science and Technology (Guangzhou), China\\
  }
\thanks{
This work was supported by Beijing University of Posts and Telecommunications-China Mobile Research Institute Joint Innovation Center. (Corresponding author: Yupeng Li, liyupengtx@126.com)}
}

\maketitle

\begin{abstract}
This letter introduces a dual application of denoising diffusion probabilistic model (DDPM)-based channel estimation algorithm integrating data denoising and augmentation. Denoising addresses the severe noise in raw signals at pilot locations, which can impair channel estimation accuracy. An unsupervised structure is proposed to clean field data without prior knowledge of pure channel information. Data augmentation is crucial due to the data-intensive nature of training deep learning (DL) networks for channel state information (CSI) estimation. The network generates new channel data by adjusting reverse steps, enriching the training dataset. To manage varying signal-to-noise ratios (SNRs) in communication data, a piecewise forward strategy is proposed to enhance the DDPM convergence precision. The link-level simulations indicate that the proposed scheme achieves a superior tradeoff between precision and computational cost compared to existing benchmarks.
\end{abstract}


\section{Introduction}
\IEEEPARstart{I}{n} massive multi-input multi-output (MIMO) systems, channel state information (CSI) plays a crucial role in optimizing receiver performance. Generally, CSI is estimated through the reference signals, predefined by both the transmitter and receiver. The precision of channel estimation significantly affects subsequent tasks like equalization and demodulation \cite{6G}. Traditional pilot-based methods, including least square (LS) and linear minimum mean square error (LMMSE), are commonly employed \cite{lmmse}.


Recently, deep learning (DL) algorithms have been integrated into channel estimation methodologies to better capture the complex propagation characteristics of wireless channels \cite{survey}. For example, some approaches frame channel interpolation and channel denoising as supervised learning problems, utilizing techniques like super-resolution convolutional neural networks (SRCNN) or image restoration (IR) networks \cite{estimat}. Another approach \cite{estimat1} treats the channel matrix as an image to leverage a denoising network for channel estimation. Although DL-based methods outperform traditional techniques, they often require substantial data collection and incur high costs associated with air interface resources\cite{magazine,gc}. This is particularly evident in supervised channel interpolation tasks, where a pilot matrix is required across the entire time-frequency domain, resulting in significant resource overhead.


The denoising diffusion probabilistic model (DDPM) \cite{dm, dm1} is well-regarded for its data generation capability, which involves introducing noise in a forward process and learning to reverse it, thereby recovering the original data. This makes DDPM attractive for communication system applications like channel equalization \cite{semantic} and denoising \cite{wcl,channelnet}. However, many existing methods assume known pure channels, a condition rarely met in real-world scenarios. Additionally, communication data often displays varying signal-to-noise ratio (SNR) levels, unlike the uniform training data typical in image or text domains, which requires a different training approach. Most research focuses on either the generative or denoising aspects of DDPM, missing the potential synergy between them.

To address these challenges, a DDPM-based channel estimation algorithm is introduced that integrates data denoising and augmentation to enhance channel estimation in communication systems. Initially, the DDPM model is trained on noisy channel matrices with diverse SNR levels, enabling it to learn the noise distribution and generate realistic channel data effectively. The estimation process then proceeds in two phases. First, DDPM generates new channel data to augment the training set for CSI estimation network, improving accuracy even with limited field data. Second, the DDPM applies its noise reduction strategies to denoise the received channel data, enhancing reliability, particularly in ``low SNR regions'' where noise can severely impact accuracy. This dual-phase approach reduces the need for extensive data collection and resource overhead while significantly enhancing channel estimation quality by leveraging the complementary generative and denoising capabilities of DDPM. Simulations demonstrate that this method achieves a superior trade-off between precision and computational cost compared to existing benchmarks\footnote{The source code is available at: \url{https://github.com/fhghwericge/Diffusion-Model-for-Enhanced-CSI-Recovery}}. 

The key contributions of this paper are as follows: 

\begin{itemize}
\item 
A novel DDPM-based channel estimation algorithm is presented that integrates data denoising and augmentation, representing a significant advancement in leveraging DDPM to reduce data collection costs while enhancing data quality.

\item 
This approach does not rely on any pure channel data throughout the process, providing a more practical alternative to many existing supervised denoising methods. By learning the noise distribution at low SNR levels, the DDPM effectively removes noise in high SNR regions, capitalizing on its generalization capabilities regarding SNR.
\item 
Considering that the received channel data often contains mixed noise levels, a piecewise forward strategy for model training is proposed, specifically designed to accommodate the unique characteristics of communication data.
\end{itemize}

\section{Configuration for DDPM}
In this section, the DDPM is employed for simultaneous field data denoising and new data generation. Initially, the DDPM is trained using field data, enabling it to effectively denoise this data. Once trained, the model can also generate augmented data by introducing noise, thereby expanding the training dataset for downstream DL tasks.

\subsection{Overview of DDPM}
For a data distribution $\textbf{H} = \textbf{H}_0 \sim p(\textbf{H}_0)$, the DDPM's forward process generates latent variables $\textbf{H}_{T}$ by adding noise at each timestep $t$ using the hyperparameter $\alpha_t \in (0,1)$ \cite{dm}, as follows:
\begin{align}
\textbf{H}_t &= \sqrt{\alpha_t}\textbf{H}_{t-1} + \sqrt{1- \alpha_t}\bm{\epsilon}_{t-1}\notag\\
&= \sqrt{\bar{\alpha}_t}\textbf{H}_{0} + \sqrt{1- \bar{\alpha}_t}\bm{\epsilon}_{0},\label{h_t}
\end{align}
where $\bm{\epsilon}_{i} \sim N_{\mathbb{C}}(\textbf{0}, \textbf{I})$, and $\bar{\alpha}_t = \prod_{i=1}^t \alpha_i$, $1\leq t\leq T$. The reverse process, like the forward process, forms a Markov chain with parameterized Gaussian transitions:
\begin{align}
\bm{p}_{\theta}(\textbf{H}_{t-1}|\textbf{H}_t) = N_{\mathbb{C}}(\textbf{H}_{t-1};\bm{\mu}_{\theta}(\textbf{H}_t,t),\sigma_t^2\textbf{I}).\label{trans}
\end{align}

The intractability of this transition is addressed by learning through a neural network (NN) via variational inference, using the tractable forward posteriors conditioned on $\textbf{H}_0$ \cite{unsupervised}. These are expressed as: $q(\textbf{H}_{t-1}|\textbf{H}_t, \textbf{H}_0) = N_{\mathbb{C}}(\textbf{H}_{t-1};\tilde{\mu}\textbf{H}_t,\textbf{H}_0),\sigma_t^2\textbf{I})$, $\tilde{\bm{\mu}}(\textbf{H}_t,\textbf{H}_0) = \frac{\sqrt{\bar{\alpha}_{t-1}(1-\alpha_t)}}{1-\bar{\alpha}_t}\textbf{H}_0 + \frac{\sqrt{\alpha_t}(1-\bar{\alpha}_{t-1})}{1-\bar{\alpha}_t}\textbf{H}_t$, and $\sigma_t^2 = \frac{(1 - \alpha_t)(1-\bar{\alpha}_{t-1})}{1-\bar{\alpha}_{t}}$. Since $\sigma_t^2$ is a time-dependent constant, the NN is trained to parameterize the conditional mean $\bm{\mu}_{\theta}(\textbf{H}_t,t)$. 

The NN function is denoted as $f_{\theta,t}^{(T)}(\textbf{H}_t) \coloneqq \bm{\mu}_{\theta}(\textbf{H}_t,t)$. Training of the DDPM is conducted by maximizing the evidence lower bound on the log-likelihood log $p(\textbf{H}_0)$ \cite{dm}. The DDPM steps can be interpreted as different SNR steps, with the SNR for step $t$ defined as:
\begin{align}
\text{SNR}_{\text{DDPM}}(t) = \frac{\mathbb{E}\left[ \left\| \sqrt{\bar{\alpha}_t \textbf{H}_0} \right\|_2^2 \right]}{\mathbb{E}\left[ \left\| \sqrt{1-\bar{\alpha}_t }\bm{\epsilon}_0 \right\|_2^2 \right]} = \frac{\bar{\alpha}_t}{1 - \bar{\alpha}_t},\label{SNR}
\end{align}
which decreases monotonically as $t$ increases.

\subsection{Noise setting strategy in DDPM}
In practical communication system, UE only receives noisy signals, and the pure channel is inaccessible. This makes channel denoising with a DDPM challenging.  To denoise without pure channel knowledge, \textbf{the noise added in the ``low SNR region'' is matched to the Gaussian noise of the received signals (i.e., in the ``high SNR region'').} Specifically, The initial step is $0$, the LS estimated channel matrix is at step $T_d$, and the DDPM's maximum step is $T_n$. The range $[0, T_d]$ can be treated as ``high SNR region'', and $[T_d, T_n]$ is the ``low SNR region''. By learning this noise distribution within the ``low SNR region'', the DDPM can effectively remove noise in the ``high SNR region''. The training phase is shown in Table \ref{denoise}, with $\textbf{H}_{T_0}$ as the precise channel, $\textbf{H}_{T_d}$ as the LS estimate, and $\textbf{H}_{T_n}$ as noisy channel, where $0<T_d<T_n$.



Once trained, the model can denoise the field data. In typical denoising DDPM, noise power is non-uniformly distributed across steps, complicating the evaluation of how many steps to reverse for effective denoising. From (\ref{h_t}), the noise power at step $t$ is $1-\bar{\alpha}_t$. To ensure uniform noise distribution, the noise power difference $Z = (1-\bar{\alpha}_t) - (1-\bar{\alpha}_{t-1})$ between two adjacent steps should be constant for any $t$. With starting value $\bar{\alpha}_1$ and ending value $\bar{\alpha}_T$, the noise power at step $t$ can be expressed as:
\begin{align}
Z_t = 1 - \bar{\alpha}_t = \frac{(1-\bar{\alpha}_{T_d)} - (1-\bar{\alpha}_1)}{T_d-1}.
\end{align}
This approach ensures a constant noise power difference across successive steps. 

Given $P_\text{signal} + P_\text{noise}=1$ and $\text{SNR} = \frac{P_\text{signal}}{P_\text{noise}}$, we find $P_{\text{noise}} = \frac{1}{1+\text{SNR}}$. The required steps for channel denoising are calculated by dividing the received noise power $P_{\text{noise}}$ by the noise power $Z_t$ per step:
\begin{align}
T_d = \frac{P_{\text{noise}}}{Z_t} = \frac{P_{\text{noise}}(T-1)}{\bar{\alpha}_1 -\bar{\alpha}_T}.\label{step_d}
\end{align}
In typical DDPM generation tasks, the goal is to generate data according to the training dataset, equating to a reverse step of $T_n - T_d$. In contrast, our work requires pure channel data for downstream tasks, so the reverse step is $T_n$ during the inference stage.

To this end, the DDPM is trained on the ``low SNR region'', while inference can be performed in both the ``low SNR region'' and the ``high SNR region''. Specifically, data denoise can be executed using the network in the ``high SNR region'', reversing from step $T_d$ to 0. Both the ``low SNR region'' and the ``high SNR region'' networks are used to generate new data, reversing from step $T_n$ to 0. This structure allows for simultaneous denoising and data augmentation. Table \ref{denoise} illustrates the inference process of DDPM for both denoising and generation. The combination of denoised field data and newly generated data forms the mixture dataset used for the DL training of downstream tasks.

\section{DDPM-based CSI learning network}
\subsection{Initial linear channel estimation}
In orthogonal frequency division multiplexing (OFDM) systems, signals are divided into frequency-domain subcarriers, which can be distorted by the wireless channel. Channel estimation is essential for accurately recovering the transmitted signals at the receiver, using pilot symbols known to both the transmitter and receiver. For the $k$-th time slot and the $i$-th subcarrier, the relationship is:
\begin{equation}
Y_{i,k} = H_{i,k}X_{i,k} + N_{i,k},
\end{equation}
where $Y_{i,k}$, $X_{i,k}$, and $N_{i,k}$ are the received signal, transmitted signal, and noise, respectively. Assume the channel response, $\bm{H}$, spans the time-frequency domain, with $H_{i,k}$ as its element at $(i,k)$, and the pilot matrix constitutes a unitary matrix. 

In LS estimation \cite{lmmse} with a unitary pilot matrix, the channel matrix where pilots reside is estimated as:
\begin{equation}
\hat{\textbf{H}}_{p}^{\text{LS}} = {\textbf{X}_p}^{-1}\textbf{Y}_p  = \textbf{H}_p + \tilde{\textbf{N}} ,
\end{equation}
where $\tilde{\textbf{N}} = {\textbf{X}_p}^{-1}\textbf{N}$ is noise with variance $\eta^2$. The LS estimate is normalized as $\textbf{H}_{p}^{\text{LS}} = (1+\eta^2)^{ -\frac{1}{2}}\hat{\textbf{H}}_{p}^{\text{LS}}$. However, LS can be noisy, especially at low SNR. The LMMSE  estimator improves upon LS by applying a filtering matrix $\textbf{W}$\cite{lmmse}, giving rise to:
\begin{equation}
\hat{\textbf{H}}_{\text{LMMSE}} = \textbf{W} \textbf{H}_{p}^{\text{LS}} = \textbf{R}_{\textbf{H}_p\textbf{H}_p} (\textbf{R}_{\text{H}_p\text{H}_p}+\frac{\beta}{\text{snr}}\textbf{I})^{-1} \textbf{H}_{p}^{\text{LS}}, 
\end{equation}
where $\textbf{R}_{\textbf{H}_p\textbf{H}_p}$ is the autocovariance matrix of  pilot matrix $\textbf{H}_p$. LMMSE effectively reduces noise but requires knowledge of $\textbf{R}_{\textbf{H}_p\textbf{H}_p}$, which can be computationally demanding in practice. 


\begin{figure}[ht]
\centering
\includegraphics[width=3.5in]{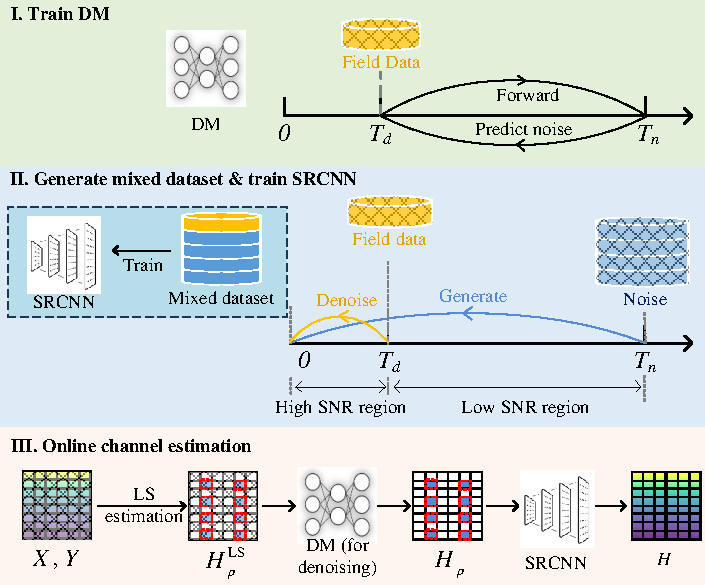}
\caption{The structure of the proposed denoising structure.} \label{inference}
\end{figure}
\subsection{Post channel interpolation}

After channel estimation, DL-based channel interpolation is commonly conducted to obtain the channel response across the entire time-frequency domain. For example, Soltani \cite{channelnet} proposed a DL-based channel estimation method using an SRCNN, treating the channel matrix as a single image. In the SRCNN network \cite{srcnn}, low-resolution (LR) images represent the channel response at pilot positions, while high-resolution (HR) images represent the complete channel response, denoted as $\textbf{H}_{\text{LR}}$ and $\textbf{H}_{\text{HR}}$, respectively. Both LR and HR images are denoised channel matrices derived from $\hat{\textbf{H}}_{\text{LS}}$. 

Without loss of generality, we adopt SRCNN for the post channel interpolation. By doing so, the SRCNN takes the channel matrix at pilot positions as input and outputs the channel matrix for the entire time-frequency domain. Since SRCNN operates as a supervised learning task, pilots must be allocated throughout the entire time-frequency domain during dataset collection, which can be resource-intensive. The LR image is then extracted from the complete channel matrix based on pilot locations. An overview of channel estimation and interpolation process is shown in Fig. \ref{inference}.

\subsection{Piecewise forwarding in model training}
In the typical image or text domains,  training data usually starts from step 0 because of the avaibility of the original data. However, in cellular systems, the received signals from different UEs often have varying SNRs. This variability means that channel samples from different UEs are distributed across different steps, requiring the DDPM to adapt accordingly. We categorize users into three representative groups: center users (dataset $D_1$ with SNR$_1$), regular users (dataset $D_2$ with SNR$_2$), and edge users (dataset $D_3$ with SNR$_3$), corresponding to step$_1$, step$_2$, and step$_3$, respectively, with step$_1<$step$_2<$step$_3$.

During training, the DDPM receives $h_k$ obtained from the forward process at step $k$, and outputs the predicted noise $n_{k-t}$. Because channel data are spread across different steps, $h_k$ might come from data with varying SNRs. In this situation, the model inputs are the same, but the expected outputs are two distinct noises, which could lead to convergence issues of the DDPM. To address this, we propose a ``piecewise forward'' strategy. Instead of extending the forward range from the current step to the maximum step, we define specific forward ranges: $D_1$ spans the range $[$step$_1$, step$_2]$, $D_2$ spans $[$step$_2$, step$_3]$, and $D_3$ spans $[$step$_3$, $T_n]$. This approach ensures samples at the same step are not forwarded from two different SNRs, improving the convergence precision of the DDPM.

\begin{table}[t!]
\normalsize
\caption{The structure of DDPM as denoise and generator.}\label{denoise}
\begin{center}
\begin{tabular}{lccc}
\hline
\textbf{Offline DDPM Training Phase}:\\
\qquad \textbf{Training dataset construction}: \\
\qquad 1. Add noise with a specifically designed\\  
\qquad  distribution to the estimated channel matrices $\textbf{H}_p^\text{LS}$.\\
\qquad 2. Apply the piecewise forward strategy to obtain \\
\qquad the training dataset for training DDPM.\\
\qquad \textbf{Input}: Training set and hyperparameters.\\
\qquad \textbf{Training stage}:\\
\qquad Train DDPM with the reverse steps \\
\qquad Select appropriate piece-wise forward steps.\\
\qquad \textbf{Output}: The trained DDPM.\\
\hline
\textbf{Online DDPM Inference Phase}:\\
\qquad \textbf{For the denoise of the field data}: \\
\qquad 1. Input the field data.\\
\qquad 2. Calculate reverse steps $T_d$ per (\ref{step_d}).\\
\qquad 3. Apply denoising to the field data.\\
\qquad \textbf{For the new data generation}: \\
\qquad 1. Input Gaussian noise.\\
\qquad 2. Set the needed reverse steps as $T_n$. \\ 
\qquad 3. Use noisy data to generate new channel data\\
\hline
\end{tabular}
\end{center}
\label{tab1}
\end{table}

\section{Simulation Results}
\subsection{Experiment setting}
In this experiment, a single antenna is employed at both the transmitter and the receiver. For this communication link, the complex channel matrix is represented as two distinct 2D images representing the real and imaginary parts. The channel data is generated using the 3GPP TR 38.901 channel simulator \cite{901}, with a carrier frequency set at 3.5 GHz. Each transmission slot consists of 12 physical downlink shared channel (PDSCH) symbols, 48 subcarriers, and 2 pilot symbols. In the simulation, the training dataset is composed of 30\% samples with an SNR of 15dB, 60\% samples with an SNR of 5dB, and 10\% samples with an SNR of -5dB, effectively simulating a typical SNR distribution in the cellular networks.

All the DL models are implemented using PyTorch, with the Adam optimizer employed for model training. The hyperparameters include an initial learning rate of $5\times 10^{-4}$, 1000 epochs, and a batch size of 256. The SRCNN is trained on a dataset of 3000 samples, consisting of 500 denoised field samples and 2500 generated samples. For the SRCNN, the hyperparameters are set to an initial learning rate of $5\times 10^{-4}$, 500 epochs, and a batch size of 256.

\subsection{Complexity analysis}
The online complexity of LS estimator is $\mathcal{O}(N_P)$, where $N_P$ represents the number of pilots. The complexity of LMMSE is primarily governed by the matrix inversion and the multiplication of multiple matrices, resulting in a complexity of $\mathcal{O}(max(N_P^3, N_LN_P^2)) = \mathcal{O}(N_LN_P^2)$. Here, $N_L=F_S \times F_D$ denotes the total number of elements in the channel matrix $\bm{H}$. The complexity of DDPM-based channel estimation is given by $\mathcal{O}(\hat{t} k^2 C_{\text{max}}^2 N_L)$, where $\hat{t}$ is the number of reverse steps in the denoising process, $k^2$ is the size of convolution kernel, and $C_{\text{max}}$ is the maximum number of channels in the convolutional network. 
The potential for parallelization on GPUs underscores the practical viability of DDPM. 

\subsection{Experiment results}
The simulations are conducted to assess the performance of our proposed channel estimation method, which integrates data augmentation and denoising with the proposed forward strategy for the training, referred to as ``A-D-T''. The traditional training strategy involves extending the forward range from the current step to the maximum step. Our algorithm is compared against three DL-based benchmarks, each evaluating the individual contributions of data augmentation, data denoising, and the DDPM training strategies introduced in our work, as outlined below:

\begin{itemize}
\item \textbf{a)} ``A-D'': This method combines data augmentation and denoising with a traditional training strategy.
\item \textbf{b)} ``A-T'': This method incorporates data augmentation with our proposed training strategy.
\item \textbf{c)} ``D-T'': This method involves denoising combined with our proposed training strategy.
\end{itemize}

All the aforementioned DL-based schemes use the same model structure and hyperparameters. The dataset comprises 500 field samples. The LS estimator is used as a lower bound of all estimators, as it does not perform additional noise processing. In contrast, assuming exact knowledge of the autocovariance matrix, the LMMSE estimator provides an upper bound for all estimators. All schemes are evaluated using clustered delay line (CDL) \cite{901} channel data.

\begin{figure}[ht]
\centering
\includegraphics[width=3.6in,height=2.7in]{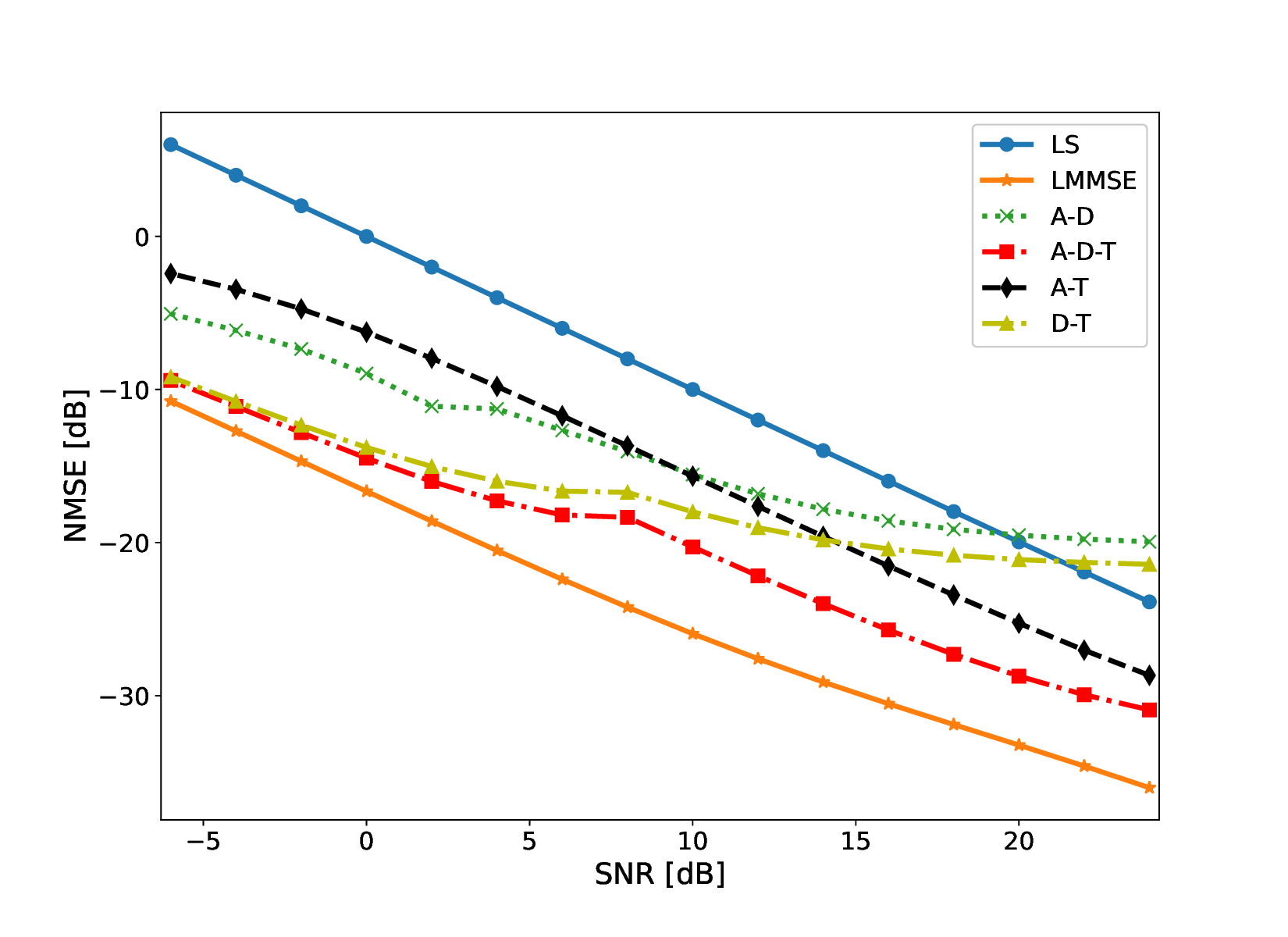}
\caption{The NMSE  comparisons among different estimation schemes.}\label{fig_SNR}
\end{figure}

Fig. \ref{fig_SNR} shows the performance for different channel estimation methods across different SNRs. The LS and LMMSE algorithms use traditional interpolation, while DDPM-based algorithms use SRCNN interpolation. LS yields the lowest precision due to noise interference. ``A-T'' improves upon LS by leveraging SRCNN's nonlinear interpolation. 

At low SNRs, ``A-D'' enhances precision over ``A-T'', demonstrating the effectiveness of data denoising. Whereas, it lags behind ``A-D-T'' due to limitations in the traditional training strategy. ``D-T'' outperforms ``A-D'' by enhancing DDPM training precision, though its performance drops at high SNRs due to insufficient field data. 

Our ``A-D-T'' method achieves the best results, especially at low SNRs, as it can reverse more steps for denoising based on the results of LS. The data with low SNR also implicitly utilizes the feature of high SNR during training. The key conclusions are: 1) Data denoising is most effective at low SNRs; 2) Data augmentation improves the performance mainly at high SNRs; 3) The training strategy improves performance across all SNRs; 4) As the training dataset is distributed only at limited SNRs, ``A-D-T'' shows strong generalization ability across diverse SNRs.


\begin{table}[ht]
\caption{The precision comparison between the mixed and pure dataset.}
\begin{center}
\begin{tabular}{p{1.6cm}p{0.6cm}p{0.6cm}p{0.6cm}p{0.6cm}p{0.6cm}p{0.6cm}}
\hline
Samples & 500 & 1000 & 1500 & 2000 & 2500 & 3000 \\
\hline
Pure dataset & -26.7 & -28.2 & -28.5 & -29.0 & -29.5 & -29.5\\
Mixed dataset & -17.5 & -25.4 & -26.0 & -27.3 & -27.9 & -27.7 \\
\hline
\end{tabular}
\label{Samples}
\end{center}
\end{table}

Table \ref{Samples} is the performance comparison between the mixed and pure datasets regarding the NMSE precision of SRCNN interpolation with SNR of 15dB. The mixed dataset includes 500 denoised field data and the rest all generated data by DDPM. The pure dataset is all pure channel data. It can be observed that, at limited samples, the pure dataset exhibits higher precision as it possesses pure data. With the increase of samples, the precision of mixed dataset approaches the pure dataset, demonstrating that the generated data has a similar quality to pure data for the training of the SRCNN, which can largely reduce the overhead of data sampling. Besides, the performance of mixed dataset increases rapidly with the increase of samples, demonstrating that data augmentation can effectively enhance the convergence precision of SRCNN.

\begin{figure}[!ht]
\centering
\includegraphics[width=3.6in,height=2.7in]{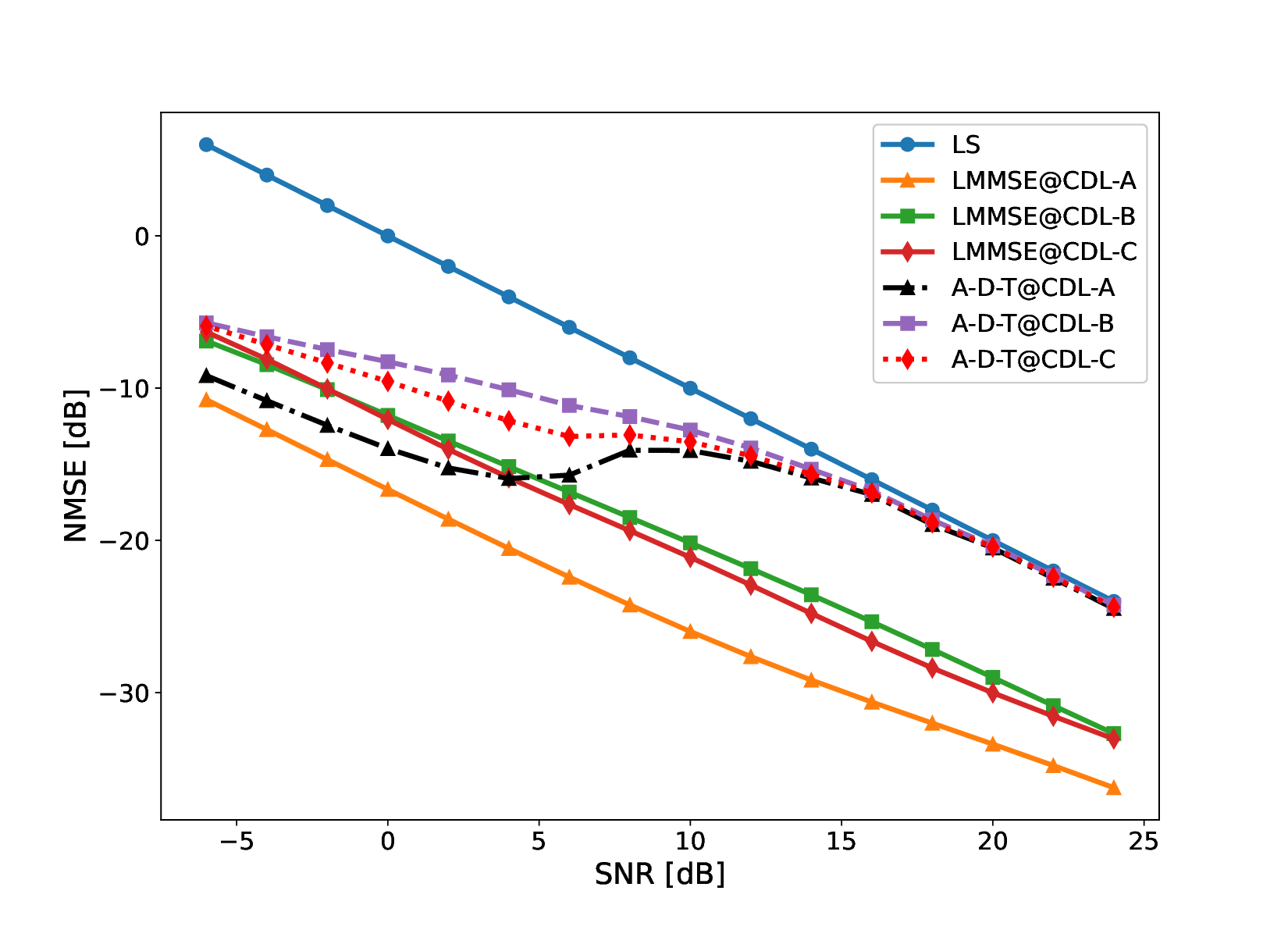}
\caption{The generalization test acrcoss other channel models.}\label{General}
\end{figure}

\begin{figure}[ht]
\centering
\includegraphics[width=3.6in,height=2.7in]{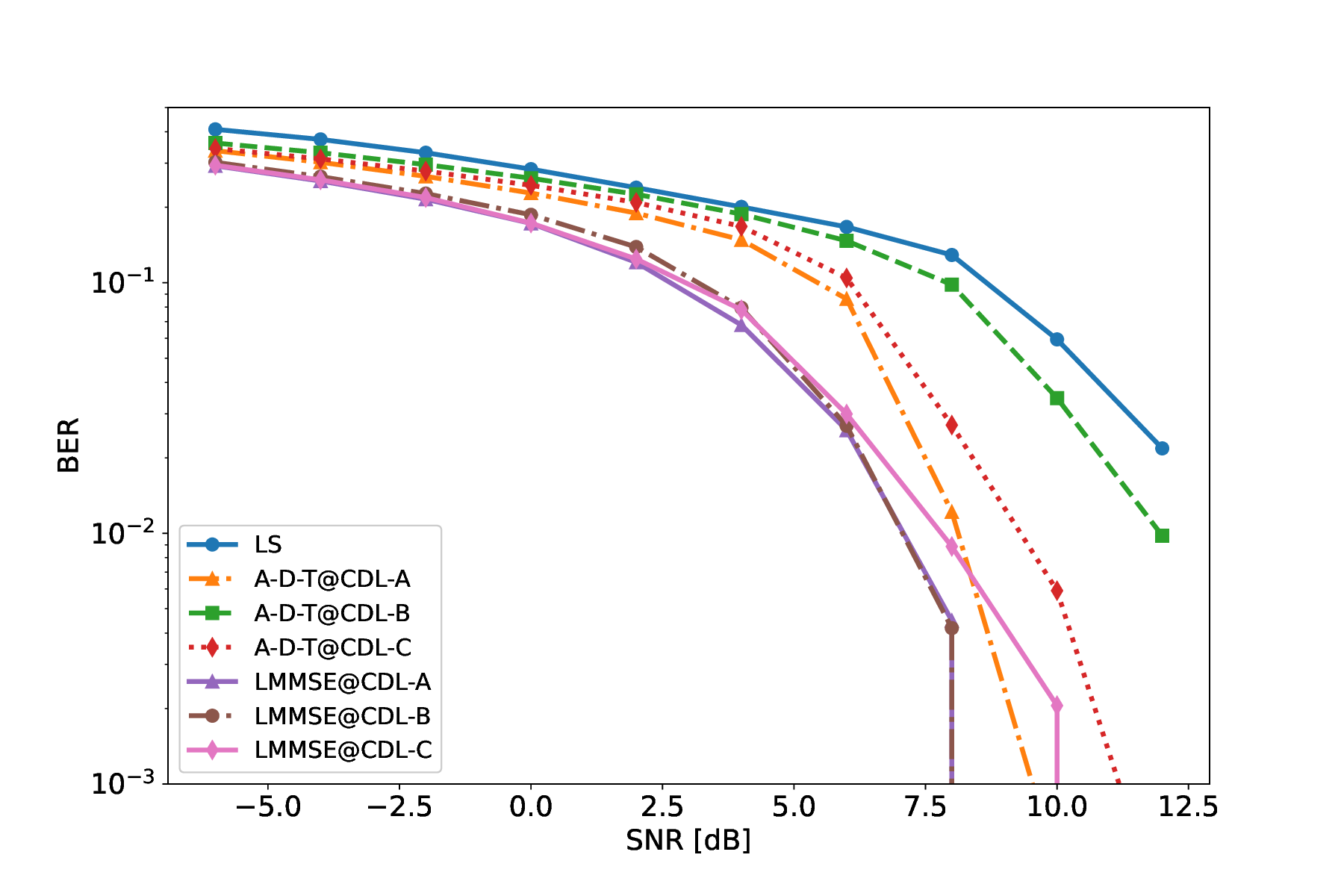}
\caption{ The BER comparisons among various schemes via link-level simulations.}\label{BER}
\end{figure}

To assess the generalization ability of DDPM, Figure \ref{General} illustrates channel estimation precision across different communication scenarios. The DDPM, trained on CDL-A channel data, is tested on CDL-A, CDL-B, and CDL-C data. The model performs best on CDL-A, its training scenario, but also maintains strong performance on CDL-B and CDL-C, with about 10 dB gain over LS and comparable precision to LMMSE at low SNR, where the DDPM can reverse more steps to accomplish denoising. A performance drop is noted after SNR reaches 5 dB. The primary reason is due to the SNR of the synthetic dataset is mostly concentrated around this region. However, we can see that, due to the DDPM learning the noise distribution features, it can effectively adapt to other channel scenarios.

For a realistic performance metric, we evaluate the end-to-end bit error rate (BER) in a digital communication chain simulation. The transmitter uses low-density parity-check (LDPC) encoding at a rate of 0.447, followed by a 64 quadrature amplitude modulation (QAM) modulation. The receiver estimates channels, performs symbol equalization, and conducts soft bit estimation and LDPC decoding. This process is repeated for 500 slots with different channel realizations. The DDPM is trained on CDL-A data, and tested on CDL-A, CDL-B, and CDL-C data. Figure \ref{BER} compares our method's end-to-end performance with LS and LMMSE estimates across different channel environments. It can be observed that our model outperforms LS in most cases. Besides, the method is never trained or tuned on any channel realization from the distribution of CDL-B and CDL-C and offers strong end-to-end performance, demonstrating the preference generalization ability of the method. 

\section{Conclusions}
This study introduces a novel DDPM-based channel estimation algorithm that enhances communication systems by integrating data augmentation and denoising. By training on noisy channel matrices across diverse SNR levels, the model effectively learns noise distribution and generates realistic channel data, improving CSI estimation accuracy while reducing data collection needs. The algorithm's advanced noise reduction strategies enhance reliability, particularly in low SNR environments. Simulations showed a remarkable gain over the traditional LS method and comparable precision to LMMSE, especially in low SNR regions, highlighting a superior balance of precision and computational efficiency. The approach generalizes well across different SNRs and scenarios, underscoring its real-world application potential. In summary, this integration within the DDPM framework significantly advances channel estimation methodologies.

\section*{Acknowledgment}
This work was funded by Beijing University of Posts and Telecommunications-China Mobile Research Institute Joint Innovation Center. The authors would like to thank Xinyu Ning for his dicussions on this article. Corresponding author: Yupeng Li, liyupengtx@126.com.

\vspace{12pt}

\end{document}